# Ligand-field helical luminescence in a 2D ferromagnetic insulator

Kyle L. Seyler[1], Ding Zhong[1], Dahlia R. Klein[2], Shiyuan Gao[3], Xiaoou Zhang[4], Bevin Huang[1], Efrén Navarro-Moratalla[2], Li Yang[3], David H. Cobden[1], Michael A. McGuire[5], Wang Yao[6], Di Xiao[4], Pablo Jarillo-Herrero[2*], Xiaodong Xu[1,7]*

[1]Department of Physics, University of Washington, Seattle, Washington 98195, USA
[2]Department of Physics, Massachusetts Institute of Technology, Cambridge, Massachusetts 02139, USA
[3]Department of Physics, Washington University, St. Louis, Missouri 63130, USA
[4]Department of Physics, Carnegie Mellon University, Pittsburgh, Pennsylvania 15213, USA
[5]Materials Science and Technology Division, Oak Ridge National Laboratory, Oak Ridge, Tennessee, 37831, USA
[6]Department of Physics and Center of Theoretical and Computational Physics, University of Hong Kong, Hong Kong, China
[7]Department of Materials Science and Engineering, University of Washington, Seattle, Washington 98195, USA

*Correspondence to: xuxd@uw.edu, pjarillo@mit.edu

**Abstract: Bulk chromium triiodide ($CrI_3$) has long been known as a layered van der Waals ferromagnet[1]. However, its monolayer form was only recently isolated and confirmed to be a truly two-dimensional (2D) ferromagnet[2], providing a new platform for investigating light-matter interactions and magneto-optical phenomena in the atomically thin limit. Here, we report spontaneous circularly polarized photoluminescence in monolayer $CrI_3$ under linearly polarized excitation, with helicity determined by the monolayer magnetization direction. In contrast, the bilayer $CrI_3$ photoluminescence exhibits vanishing circular polarization, supporting the recently uncovered anomalous antiferromagnetic interlayer coupling in $CrI_3$ bilayers[2]. Distinct from the Wannier-Mott excitons that dominate the optical response in well-known 2D van der Waals semiconductors[3], our absorption and layer-dependent photoluminescence measurements reveal the importance of ligand-field and charge-transfer transitions to the optoelectronic response of atomically thin $CrI_3$. We attribute the photoluminescence to a parity-forbidden *d-d* transition characteristic of $Cr^{3+}$ complexes, which displays broad linewidth due to strong vibronic coupling and thickness-independent peak energy due to its localized molecular orbital nature.**

**Main Text:**

Van der Waals layered materials offer fascinating opportunities for studying light-matter interactions in the 2D limit. For instance, monolayer semiconducting transition metal dichalcogenides (e.g. $WSe_2$) enable coupling between the helicity of light and the valley degree of freedom[4]. In all non-metallic 2D materials to date, it has been established that tightly bound Wannier-Mott excitons dominate the intrinsic optical response[3], and there has been rapid progress in studying 2D excitonic interactions, dynamics, and spin/valley physics[3,5]. However, none of these 2D materials possess long-range magnetic order. A monolayer semiconductor or insulator with intrinsic magnetism would enable the study of novel photo-physical phenomena and the interplay with underlying magnetic order, possibly involving physics incompatible with the Wannier-Mott excitonic picture.

On the other hand, the exploration of ferromagnetism in non-metallic bulk materials has a long history. Early studies examined the intrinsic ferromagnetic ordering of a variety of insulating and semiconducting materials, including for example, the ferrites and ferrospinels[6], Cr trihalides[7], Eu chalcogenides[8], and Cr spinels[9]. Later, with the introduction of magnetic dopants into nonmagnetic II-VI and III-V semiconductors, diluted magnetic semiconductors captured widespread attention[10], boosted by the discovery of ferromagnetism in Mn-doped InAs (ref. 11) and GaAs (ref. 12) in the 1990s. Central to progress in these fields, optical experiments have led to deep understanding of electronic structure, magnetization dynamics, and interactions between magnetism and light[8,13–15]. While the fascinating physics in the quantum structures of diluted magnetic semiconductors has propelled spintronics research over the last few decades[16], there has been a comparative lack of non-metallic nanoscale materials hosting *intrinsic* magnetism.

Recently, a number of van der Waals magnetic insulators have emerged as a promising platform for exploring light-matter interactions in the monolayer limit[2,17,18]. In particular, a magneto-optical Kerr effect (MOKE) study[2] revealed 2D ferromagnetism in monolayer $CrI_3$ (Fig. 1a). Historically, the bulk chromium trihalides were found to behave as Mott insulators with an optical response governed by ligand-field and charge-transfer transitions[13,19–22], which are highly localized photo-excitations between molecular orbitals. Therefore, atomically thin chromium trihalides may provide a new platform from which to study 2D optical physics under the influence of intrinsic magnetic ordering beyond the excitonic picture. In this work, we reveal magnetization-determined ligand-field transitions in 2D ferromagnetic $CrI_3$ by magneto-photoluminescence (PL) and reflection spectroscopy.

We prepared monolayer $CrI_3$ on sapphire substrates by mechanical exfoliation of bulk crystals. For the PL experiments, we excited with a linearly polarized laser at 1.96 eV and analyzed the circularly polarized PL components (see Methods). All measurements were done under normal incidence excitation and collection. We first measured the monolayer $CrI_3$ PL at 75 K, above the Curie temperature ($T_C$) of 45 K (ref. 2), without applying magnetic field. As shown in Fig. 1b, there is a single broad emission feature centered at ~1.1 eV with a full-width at half-maximum of about 180 meV. We observed no other PL features at higher energy. The red and blue spectra correspond to $\sigma^+$ and $\sigma^-$ circularly polarized PL, respectively. The two components are indistinguishable, which is consistent with the absence of magnetic order above $T_C$.

Remarkably, upon cooling to 15 K (well below $T_C$) in the absence of a magnetic field, we observe that the $\sigma^+$ and $\sigma^-$ PL diverge in intensity, with $\sigma^+$ much brighter than $\sigma^-$ (Fig. 1c). Defining the PL circular polarization, $\rho = \frac{I_+ - I_-}{I_+ + I_-}$, where $I_\pm$ is the peak intensity of $\sigma^\pm$ PL, we find $\rho \approx 0.45$. Since the linearly polarized excitation does not break the time reversal symmetry of the system, the observation of net circular emission demonstrates the appearance of spontaneous out-of-plane magnetic ordering in monolayer $CrI_3$. While the peak narrows slightly (~10%) on cooling from 75 K to 15 K, its position does not change. Any energy splitting between the two polarizations is smaller than ~1 meV and unresolvable. The temperature dependence of $\rho$ is shown in Fig. 1e. The onset of finite $\rho$ at 45 K is consistent with $T_C$ for a monolayer[2].

We next explore the effect of an applied out-of-plane magnetic field. Fig. 2a shows the circularly polarized PL spectra at +0.5 T. Here the polarization is reversed relative to that shown in Fig. 1c (which was taken at zero field), due to a flip in the magnetization by the applied field. If the field is then lowered to 0 T, the spectra remain unchanged (Fig. 2b). When the opposite field, -0.5 T, is then applied, the magnetization is reversed again and the PL returns to its original state with a stronger $\sigma^+$ component than $\sigma^-$. This state is in turn preserved when the field is returned to zero (Figs. 2c, d). Additional data demonstrating this behavior for a monolayer on a $SiO_2$/Si substrate is presented in Supplementary Section 1. We show ρ over a cycle of the magnetic field in Fig. 2e, where the observed hysteresis loop is clearly the hallmark of ferromagnetic behavior. The saturation polarization is ±0.5, and the coercive field is ~55 mT. Both circular polarization components show linear power dependence (Supplementary Section 2), leading to a power-independent ρ.

The field-dependent measurements unambiguously demonstrate that the helicity of the $CrI_3$ PL is determined by its magnetic ordering, and thus that measurement of PL helicity can be used as a simple probe of its magnetic phases. Figures 3a-c present the circular polarization-resolved PL from a bilayer of $CrI_3$ at 15 K, at magnetic fields -1, 0, and +1 T respectively. We see strong circularly polarized PL at ±1 T (Figs. 3a, c), consistent with full spin polarization in bilayer $CrI_3$ at such high fields. In contrast with the monolayer, no net circular polarization is seen at zero field (Fig. 3b), which implies vanishing net out-of-plane magnetization. As shown in Fig. 3d, the polarization ρ is negligible between ±0.65 T and jumps abruptly to a value of -0.5 above 0.65 T and +0.5 below -0.65 T. This magnetic-field-dependent PL polarization forms a close parallel to the MOKE response in Ref. 2, which also vanishes in the same field range. The consistency of the results for these two distinct experimental probes reinforces the interpretation that at low fields, bilayer $CrI_3$ consists of two ferromagnetic monolayers that are antiferromagnetically coupled.

What is the origin of luminescence in atomically thin $CrI_3$? In well-studied 2D semiconductors, such as $WSe_2$, the photo-response can be described by band-to-band transitions with strong excitonic effects. Tightly bound Wannier-Mott excitons dominate the optical spectrum[3]. In $CrI_3$, however, the 3$d$ electrons are much more spatially localized, and thus to understand the optical response, a molecular orbital approach is more suitable. In fact, ligand field theory, which predicts intra-atomic *d-d* transitions and higher energy charge-transfer transitions within metal-ligand complexes[23], has previously been used to interpret the optical spectra of bulk transition metal halides (refs 13,19–22). While prior studies have focused on absorption/reflection in $CrCl_3$ and $CrBr_3$, there are few reports on the optical properties of $CrI_3$ (refs 1,13,19), and none examining the PL. Furthermore, all previous work was limited to bulk crystals containing thousands of layers. We now present evidence that these intrinsic localized photo-excitations dominate in 2D $CrI_3$.

We first reiterate that the monolayer PL intensity is linear in excitation power, as illustrated in Supplementary Fig. 2. Taken together with the tight link between the PL helicity and layer-dependent magnetic phases, this rules out the possibility that the dominant PL contribution is from defect-bound excitons, which tend towards saturation at higher excitation intensity and are often seen in the low-energy PL spectra of other 2D semiconductors[3,24]. To investigate the electronic response more broadly, we measured the differential reflectance of monolayer $CrI_3$ on sapphire (Fig. 4a), which is proportional to its absorbance[25]. There is a weak peak near 1.5 eV along with stronger features around 2 eV and 2.7 eV. Using the ligand-field framework, we can attribute the 2.7 eV peak as well as the strongest two peaks near 2 eV to dipole-allowed ligand-to-metal charge-transfer (LMCT) transitions between the iodine 5$p$ orbitals and $Cr^{3+}$ 3$d$ orbitals (Supplementary

Sections 3 and 4). The 1.5 eV transition has not been discussed much before in the literature. Because of the approximate octahedral symmetry of the iodine ligands around each $Cr^{3+}$ site (Fig. 1a), the $d^3$ configuration ($^4F$ term) in isolated $Cr^{3+}$ splits into an $^4A_2$ ground state and $^4T_2$ and $^4T_1$ excited states of the $t_{2g}$ and $e_g$ orbitals in $CrI_3$. We assign the 1.5 eV peak to the lowest energy transition between these levels, from $^4A_2$ to $^4T_2$. Despite being electric-dipole forbidden by the Laporte parity selection rule, $d$-$d$ transitions can become weakly allowed by mixing with odd parity states, such as produced by phonons[26]. In addition, the trigonal field of the nearest-neighbor Cr atoms (Fig. 1a) eliminates the local inversion symmetry of each Cr site, providing a static odd-parity field to allow the $d$-$d$ transitions. We deduce that the absorbance of the $^4T_2$ transition is about 0.7% for monolayer $CrI_3$ (see Methods), which is an order of magnitude lower than the absorbance for the A and B excitonic resonances in monolayer semiconducting transition metal dichalcogenides[27], such as $MoS_2$. The weak absorption underscores the weakly allowed parity-forbidden nature of the $d$-$d$ transition.

Our assignment of the reflection features is consistent with the results of prior experiments and recent calculations[28,29] on bulk $CrCl_3$ and $CrBr_3$. To connect these with $CrI_3$, in Fig. 4b, we plot the absorption peak energies of bulk $CrI_3$ that we measured (Supplementary Section 4) against those of $CrCl_3$ and $CrBr_3$ from previous studies[13,19–22]. The relationship between the optical spectra of the Cr trihalides then becomes clear. The large energetic shift of the two high-energy peaks between different ligand species confirms their charge transfer origin. On the other hand, the $d$-$d$ transitions exhibit weaker dependence on the ligand. Interestingly, in view of the decreased ligand-field strength of $I^-$ in $CrI_3$, we expect the lowest LMCT transition to overlap the highest $d$-$d$ transition ($^4A_2$ to $^4T_1$). This suggests that the origin of the low-energy shoulder at 1.8 eV is absorption to the $^4T_1$ level, enhanced by its energetic proximity to the strong charge transfer transitions near 2 eV. Thus, it is likely that both LMCT and $d$-$d$ transitions play a role in the recently observed magneto-optical Kerr response[2] near 2 eV in few-layer $CrI_3$.

We can now begin to understand the origin of PL in atomically thin $CrI_3$. The clear correlation between the three Cr trihalide compounds in the energies of the lowest absorption peak and PL (Fig. 4b) implies that the monolayer $CrI_3$ PL arises from the $^4T_2$ to $^4A_2$ $d$-$d$ transition. The ~400 meV Stokes shift between the PL and 1.5 eV reflection peak is a consequence of the Franck-Condon principle and strong electron-lattice coupling[26] (Fig. 4c, Huang-Rhys factor ~ 10, see Supplementary Section 5). The extracted ~1.3 eV $^4T_2$ zero-phonon energy is also consistent with the estimated ligand-field splitting of 1.2 eV determined from the angular overlap between the Cr-I $\sigma$ and $\pi$ bonds (Supplementary Section 5). The wide PL linewidth is a signature of $d$-$d$ luminescence broadened by vibronic modes, characteristic of transition metal ions in solids[26]. In addition, our layer-dependent study demonstrates that the PL peak energy is independent of the thickness of $CrI_3$ (Fig. 4d, e), highlighting the localized nature of the ligand-field excitation. Because the $d$-$d$ transition is parity-forbidden, the observed PL intensity will sensitively depend on the origins of local symmetry breaking, such as from odd-parity phonons and trigonal distortion, and on the interactions with excited states and spin-orbit coupling. Determining the strengths of these interactions will be essential for understanding the relative intensity of the PL helicities observed in ferromagnetic $CrI_3$. We also note that PL intensity per layer (i.e. normalized by the number of layers) increases with increasing thickness (Fig. 4e), suggesting that the $d$-$d$ relaxation processes are affected by interlayer or substrate interactions.

The above measurements establish the prominence of highly localized optical excitations in atomically thin $CrI_3$. While transport measurements will be necessary to confirm its electrical properties, mono- and few-layer CrI$_3$ are likely to be good insulators, judging from the localized optical response and the large bulk resistivity[30]. The ligand-field luminescence helicity displays clear signatures of the underlying magnetic order, suggesting that the optical spectra will serve as an important probe of magnetic order in atomically thin CrI$_3$. Our results enlarge the landscape of light-matter interactions in 2D materials, and suggest new opportunities to study and control ligand-field spectra in the 2D limit in the presence of magnetic ordering. Further magneto-optical studies on the thickness and polarization dependence of the charge-transfer and ligand-field transitions may shed light on the nature of the intralayer and interlayer exchange interactions. Beyond this, our work previews exciting possibilities to use CrI$_3$ as an atomically thin magnetic insulator in van der Waals heterostructures. Complementary to the popular non-magnetic layered insulator hexagonal boron nitride, 2D CrI$_3$ will serve as a substrate, interfacial layer, and tunnel barrier for engineering magnetic proximity effects, exploring spin-dependent tunneling phenomena, and designing novel magneto-optoelectronic devices with spontaneous helical light emission.

## Methods

### Sample fabrication

Bulk CrI$_3$ crystals were grown by chemical vapor transport, as described in detail in Ref. 2 and 30. Monolayer and bilayer CrI$_3$ samples were then obtained by mechanical exfoliation from bulk CrI$_3$ onto a 0.5 mm thick *c*-plane sapphire substrate in an Ar-filled glovebox. We also fabricated and measured samples on 285 nm SiO$_2$/Si (Supplementary Fig. 1 and Fig. 3d). We confirmed the optical contrast of bilayer CrI$_3$ on sapphire by transferring a bilayer from 285 nm SiO$_2$/Si (on which the optical contrast has been established[2]) to sapphire. Thus, we determined the optical contrast on sapphire to be ~0.035 and ~0.07 at 631 nm for monolayer and bilayer CrI$_3$, respectively. Samples were kept under inert atmosphere or vacuum during the entire fabrication and measurement process.

### Optical measurements

Low-temperature optical measurements were performed in a closed-cycle cryostat with a superconducting magnet with the axis directed out of the sample plane. For the photoluminescence measurements, the sample was excited by a HeNe laser (632.8 nm) focused to a ~1 µm spot diameter. A low power of 10 µW was used to avoid sample heating and degradation. A dichroic beam splitter reflected the collected PL, which was then spatially filtered through a confocal pinhole (to avoid collecting nearby bulk CrI$_3$ PL), dispersed by a 1.2 m blaze grating, and detected by a liquid-nitrogen cooled InGaAs linear photodiode array (Princeton Instruments). The InGaAs detector was spectrally calibrated using Hg emission lines. The excitation and detection polarization were controlled using linear polarizers and achromatic near-infrared half- and quarter-wave plates. Peak intensities were calculated by averaging 100 points (~30 meV) about the peak center. For the white light differential reflection measurements, we spatially filtered a tungsten halogen lamp through a pinhole and focused the beam to a ~3 µm spot size on the CrI$_3$. The reflected light was deflected with a beam splitter and detected by a spectrometer and Si CCD or InGaAs array, which enabled measurements from 1 to 3 eV. To obtain the differential reflectance, we subtracted and normalized the CrI$_3$ reflectance by the reflectance of the bare sapphire substrate.

The $^4T_2$ absorbance was determined as $\frac{1}{4}(n^2-1)\frac{\Delta R}{R}$ (ref. 25), where $n \approx 1.76$ is the ordinary refractive index of sapphire at 1.5 eV (ref. 31).

**References:**

1. Dillon, J. F. & Olson, C. E. Magnetization, resonance, and optical properties of the ferromagnet $CrI_3$. *J. Appl. Phys.* **36,** 1259–1260 (1965).
2. Huang, B. *et al.* Layer-dependent ferromagnetism in a van der Waals crystal down to the monolayer limit. *Nature* **546,** 270–273 (2017).
3. Mak, K. F. & Shan, J. Photonics and optoelectronics of 2D semiconductor transition metal dichalcogenides. *Nat. Phot.* **10,** 216–226 (2016).
4. Xu, X., Yao, W., Xiao, D. & Heinz, T. F. Spin and pseudospins in layered transition metal dichalcogenides. *Nat. Phys.* **10,** 343–350 (2014).
5. Schaibley, J. R. *et al.* Valleytronics in 2D materials. *Nat. Rev. Mater.* **1,** 16055 (2016).
6. Snoek, J. L. *New developments in ferromagnetic msaterials*. (Elsevier, 1947).
7. Tsubokawa, I. On the magnetic properties of a $CrBr_3$ single crystal. *J. Phys. Soc. Japan* **15,** 1664–1668 (1960).
8. Wachter, P. The optical electrical and magnetic properties of the europium chalcogenides and the rare earth pnictides. *C R C Crit. Rev. Solid State Sci.* **3,** 189–241 (1972).
9. Baltzer, P. K., Lehmann, H. W. & Robbins, M. Insulating ferromagnetic spinels. *Phys. Rev. Lett.* **15,** 493–495 (1965).
10. Furdyna, J. K. Diluted magnetic semiconductors. *J. Appl. Phys.* **64,** R29–R64 (1988).
11. Ohno, H., Munekata, H., Penney, T., von Molnár, S. & Chang, L. L. Magnetotransport properties of p-type (In,Mn)As diluted magnetic III-V semiconductors. *Phys. Rev. Lett.* **68,** 2664–2667 (1992).
12. Ohno, H. *et al.* (Ga,Mn)As: A new diluted magnetic semiconductor based on GaAs. *Appl. Phys. Lett.* **69,** 363–365 (1996).
13. Dillon, J. F., Kamimura, H. & Remeika, J. P. Magneto-optical properties of ferromagnetic chromium trihalides. *J. Phys. Chem. Solids* **27,** 1531–1549 (1966).
14. Burch, K. S., Awschalom, D. D. & Basov, D. N. Optical properties of III-Mn-V ferromagnetic semiconductors. *J. Magn. Magn. Mater.* **320,** 3207–3228 (2008).
15. Miles, P. A., Westphal, W. B. & Von Hippel, A. Dielectric spectroscopy of ferromagnetic semiconductors. *Rev. Mod. Phys.* **29,** 279–307 (1957).
16. Dietl, T. & Ohno, H. Dilute ferromagnetic semiconductors: Physics and spintronic structures. *Rev. Mod. Phys.* **86,** 187–251 (2014).
17. Gong, C. *et al.* Discovery of intrinsic ferromagnetism in 2D van der Waals crystals. *Nature* **546,** 265–269 (2017).
18. Lee, J.-U. *et al.* Ising-type magnetic ordering in atomically thin $FePS_3$. *Nano Lett.* **16,** 7433–7438 (2016).
19. Grant, P. M. & Street, G. B. Optical properties of the chromium trihalides in the region 1-11 eV. *Bull. Am. Phys. Soc. II* **13,** (1968).
20. Pollini, I. & Spinolo, G. Intrinsic optical properties of $CrCl_3$. *Phys. status solidi* **41,** 691–701 (1970).
21. Bermudez, V. M. & McClure, D. S. Spectroscopic studies of the two-dimensional magnetic insulators chromium trichloride and chromium tribromide—I. *J. Phys. Chem. Solids* **40,** 129–147 (1979).

22. Nosenzo, L., Samoggia, G. & Pollini, I. Effect of magnetic ordering on the optical properties of transition-metal halides: $NiCl_2$, $NiBr_2$, $CrCl_3$, and $CrBr_3$. *Phys. Rev. B* **29,** 3607–3616 (1984).
23. Figgis, B. N. & Hitchman, M. A. *Ligand field theory and its applications*. (Wiley-VCH, 2000).
24. Tongay, S. *et al.* Defects activated photoluminescence in two-dimensional semiconductors: interplay between bound, charged, and free excitons. *Sci. Rep.* **3,** 2657 (2013).
25. McIntyre, J. D. E. & Aspnes, D. E. Differential reflection spectroscopy of very thin surface films. *Surf. Sci.* **24,** 417–434 (1971).
26. Henderson, B. & Imbusch, G. F. *Optical spectroscopy of inorganic solids*. (Clarendon Press, 1989).
27. Li, Y. *et al.* Measurement of the optical dielectric function of monolayer transition-metal dichalcogenides: $MoS_2$, $MoSe_2$, $WS_2$, and $WSe_2$. *Phys. Rev. B* **90,** 205422 (2014).
28. Shinagawa, K., Sato, H., Ross, H. J., McAven, L. F. & Butler, P. H. Charge-transfer transitions in chromium trihalides. *J. Phys. Condens. Matter* **8,** 8457 (1996).
29. Butler, L. F. M. and H. J. R. and K. S. and P. H., McAven, L. F., Ross, H. J., Shinagawa, K. & Butler, P. H. The Kerr magneto-optic effect in ferromagnetic $CrBr_3$. *J. Phys. B At. Mol. Opt. Phys.* **32,** 563 (1999).
30. McGuire, M. A., Dixit, H., Cooper, V. R. & Sales, B. C. Coupling of crystal structure and magnetism in the layered, ferromagnetic insulator $CrI_3$. *Chem. Mater.* **27,** 612–620 (2015).
31. Malitson, I. H. Refraction and dispersion of synthetic sapphire. *J. Opt. Soc. Am.* **52,** 1377–1379 (1962).

**Acknowledgements:** The authors thank Daniel Gamelin for insightful discussions on the optical response of $CrI_3$, and Arka Majumdar for testing the measurement system. Work at the University of Washington was mainly supported by the Department of Energy, Basic Energy Sciences, Materials Sciences and Engineering Division (DE-SC0018171), and University of Washington Innovation Award. Work at MIT has been supported by the Center for Integrated Quantum Materials under NSF grant DMR-1231319 as well as the Gordon and Betty Moore Foundation's EPiQS Initiative through Grant GBMF4541 to PJH. Device fabrication has been partly supported by the Center for Excitonics, an Energy Frontier Research Center funded by the US Department of Energy (DOE), Office of Science, Office of Basic Energy Sciences under Award Number DESC0001088. DC's contribution is supported by DE-SC0002197. Work at CMU is also supported by DOE BES DE-SC0012509. WY is supported by the Croucher Foundation (Croucher Innovation Award), the RGC of Hong Kong (HKU17305914P), and the HKU ORA. Work at ORNL (MAM) was supported by the US Department of Energy, Office of Science, Basic Energy Sciences, Materials Sciences and Engineering Division. XX and DX acknowledge the support of a Cottrell Scholar Award. SG and LY are supported by NSF grant No. DMR-1455346 and EFRI-2DARE-1542815. XX acknowledges the support from the State of Washington funded Clean Energy Institute and from the Boeing Distinguished Professorship in Physics.

**Author Contributions:** XX, KLS, and PJH conceived the experiment. KLS built the experimental setup and carried out the measurements, assisted by DZ and BH, supervised by XX. Crystal growth, characterization, and device fabrication at MIT were carried out by DRK and ENM, supervised by PJH. Device fabrication at UW was carried out by KLS, DZ, and BH, with crystal grown and characterized by MAM at ONRL. KLS and XX analyzed and interpreted the data with theoretical support from XZ, DX, WY, SG, and LY. KLS, XX, DHC, and PJH wrote the manuscript with input from all authors. All authors discussed results.

**Competing Financial Interests:** The authors declare no competing financial interests.

**Figures:**

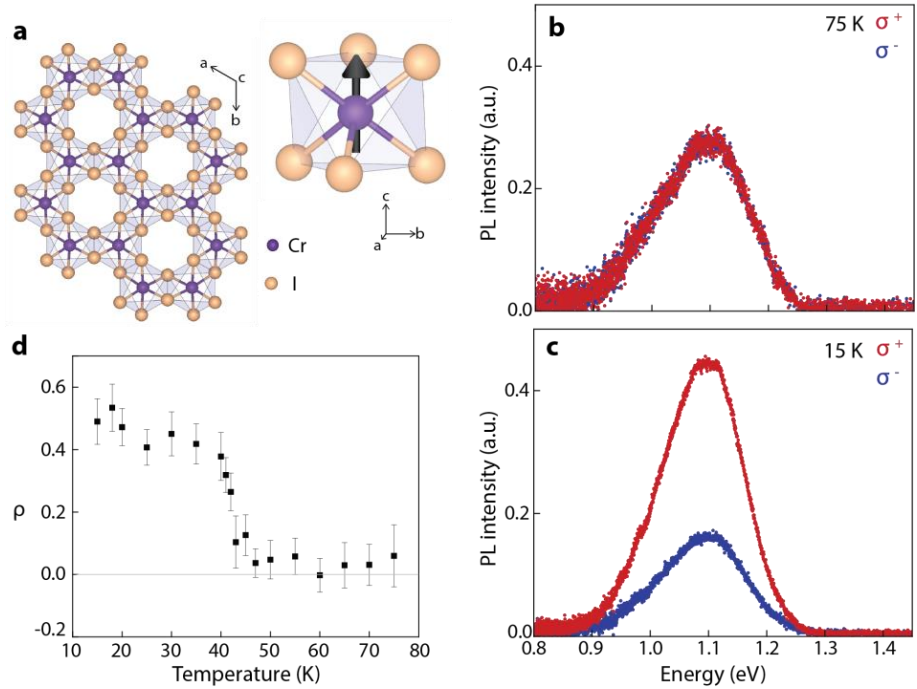

**Figure 1. Spontaneous circularly polarized luminescence from monolayer CrI$_3$.** (**a**) Top view of the crystal structure of CrI$_3$ (left) and side view (right) depicting single Cr site with an arrow representing its out-of-plane magnetic moment. (**b**) Photoluminescence (PL) spectrum for $\sigma^+$ (red) and $\sigma^-$ (blue) circularly polarized PL from a monolayer at 75 K. The excitation is linearly polarized. (**c**) Same as (b), but at 15 K. (**d**) Temperature dependence of the degree of PL circular polarization ($\rho$) at 0 T from 75 to 15 K. Error bars show the standard deviation of the polarization at the peak. No external magnetic field was applied while acquiring the data in Fig. 1b-d.

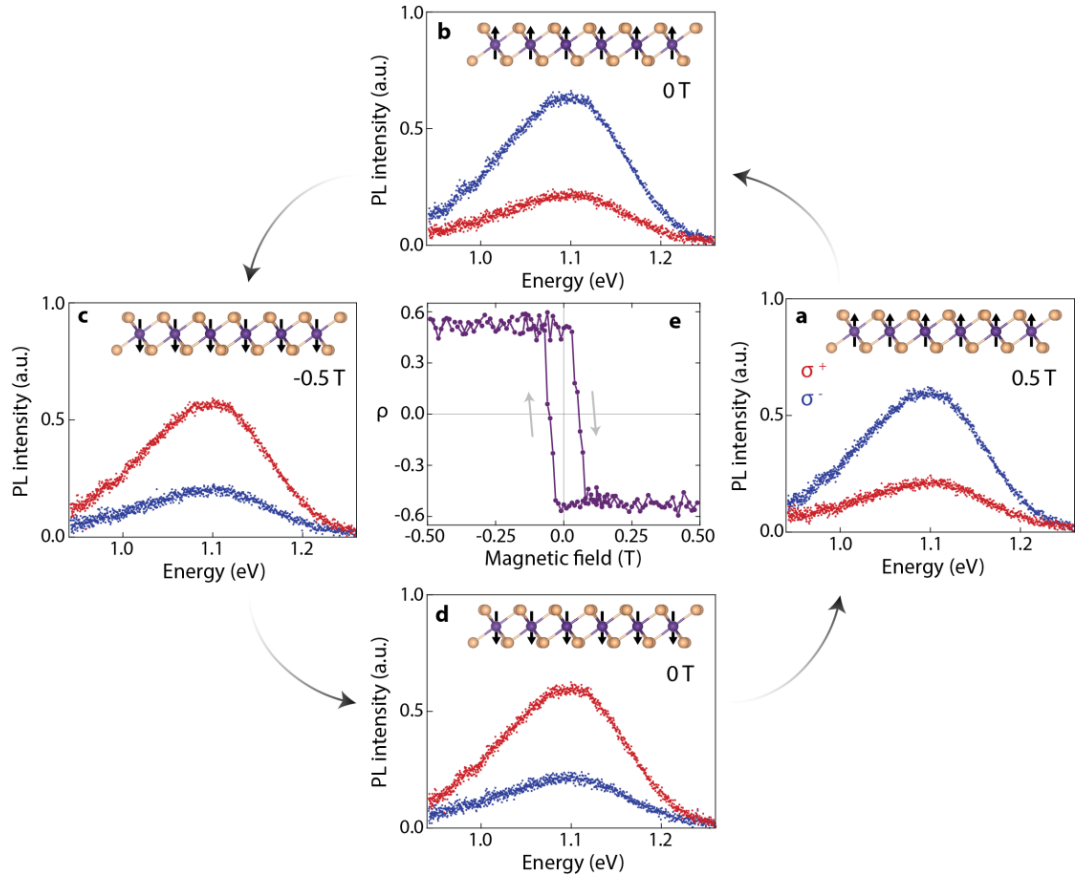

**Figure 2. Photoluminescence from monolayer CrI₃ in an applied magnetic field.** Sequence of PL spectra for $\sigma^+$ (red) and $\sigma^-$ (blue) circular polarization components, acquired at 15 K under linearly polarized excitation. **(a)** 0.5 T (increasing from 0 T), **(b)** 0 T (decreasing from 0.5 T), **(c)** -0.5 T (decreasing from 0 T), and **(d)** 0 T (increasing from -0.5 T). The curved arrows indicate the magnetic field sweep direction, and the insets depict the magnetization direction relative to the lattice. **(e)** Circular polarization (ρ) as a function of applied field over one full cycle. Gray arrows show the sweep direction of the applied field.

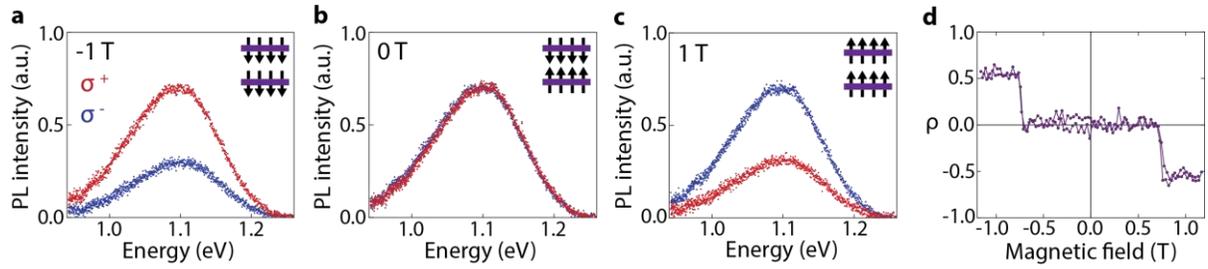

**Figure 3. Bilayer luminescence reveals antiferromagnetic ground state.** PL spectrum of a bilayer sample for $\sigma^+$ (red) and $\sigma^-$ (blue) circular polarization and linearly polarized excitation acquired at 15 K at magnetic fields of -1 T **(a)**, 0 T **(b)**, and +1 T **(c)**. The inset figures depict the inferred magnetization pattern of the bilayer. It should be noted that at 0 T the net magnetization is zero, but the exact spin orientation within each layer is unknown. **(d)** Circular polarization ($\rho$) as a function of magnetic field. Data points for increasing and decreasing magnetic field overlap to within uncertainty. The data in (a)-(c) are from a bilayer on sapphire while the data in (d) are from a bilayer on $SiO_2$. Their behavior is consistent.

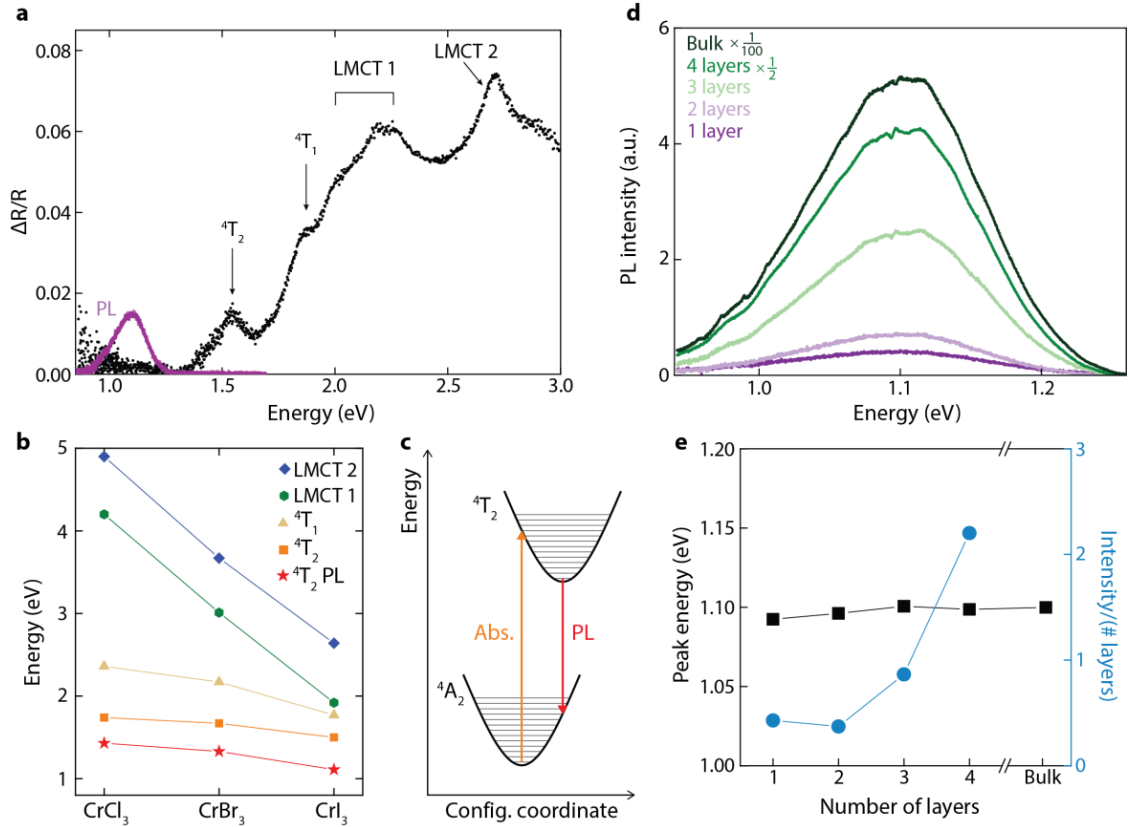

**Figure 4. Reflection spectrum and thickness-dependent PL.** (**a**) Differential reflection spectrum of monolayer $CrI_3$ (black) and overlaid PL spectrum (purple). See text for discussion of labelled peak assignments. (**b**) Energies of the corresponding peaks in bulk $CrCl_3$, $CrBr_3$, and $CrI_3$. The $CrI_3$ data are from the sample in Supplementary Fig. 4, while the data for $CrBr_3$ and $CrCl_3$ are compiled from Refs. 13,19–22. (**c**) Configurational coordinate diagram in the harmonic approximation for the observed ligand-field $^4T_2$ absorption and PL. The horizontal lines represent vibrational levels. The calculated Huang-Rhys factor is ~10. (**d**) Layer dependence of the PL spectra at 15 K and zero magnetic field. Note that the 4-layer-thick and bulk spectra have been divided by a factor of 2 and 100 respectively. The small features near 1.08 eV are due to a slight artefact of the grating which could not be corrected. (**e**) Peak energy (black) and PL intensity normalized by number of layers (blue) for different thicknesses. The peak energy is calculated by weighted average using the spectra in Fig. 4d.

# Supplementary Materials for

## Ligand-field helical luminescence in a 2D ferromagnetic insulator

Contents:

## S1. PL from monolayer CrI$_3$ on SiO$_2$/Si

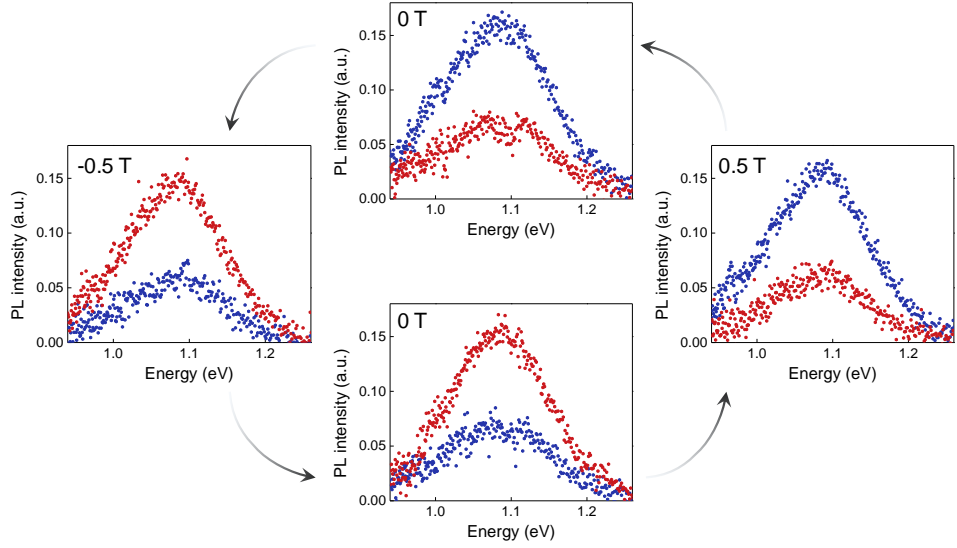

**Fig. S1.** PL spectra from monolayer CrI$_3$ on SiO$_2$/Si substrate for $\sigma^+$ (red) and $\sigma^-$ (blue) circular polarization, acquired at 15 K under linearly polarized excitation and different applied magnetic fields perpendicular to the sample plane. Magnetic field values are shown on the plots and the arrows indicate the magnetic field sweep direction. The results are consistent with those from sapphire substrate (Fig. 2). We note that PL from the underlying Si shows up near 1.1 eV. We directly subtracted this PL contribution by using the background spectrum from uncovered SiO$_2$/Si. To avoid the complication from background Si PL, it is better to study the detailed features of the CrI$_3$ PL spectrum on transparent substrates.

## S2. Power dependence of monolayer PL

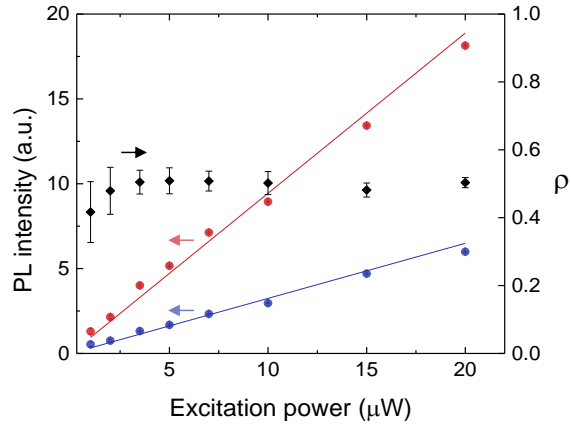

**Fig. S2.** Power dependence of $\sigma^+$ (red) and $\sigma^-$ (blue) PL peak intensities and circular polarization ($\rho$, black). The data is taken at 0 T after decreasing from -0.5 T (same condition as in Fig. 2d). Red and blue lines show a linear fit to the intensity data. We note that the low count rate prohibits accurate measurement of the PL below 1 $\mu$W. Furthermore, the power was kept less than or equal to 20 $\mu$W to avoid possible sample degradation. Error bars show the standard deviation of the intensities and $\rho$ at the peak.

## S3. Molecular orbital energy diagram of CrI$_3$

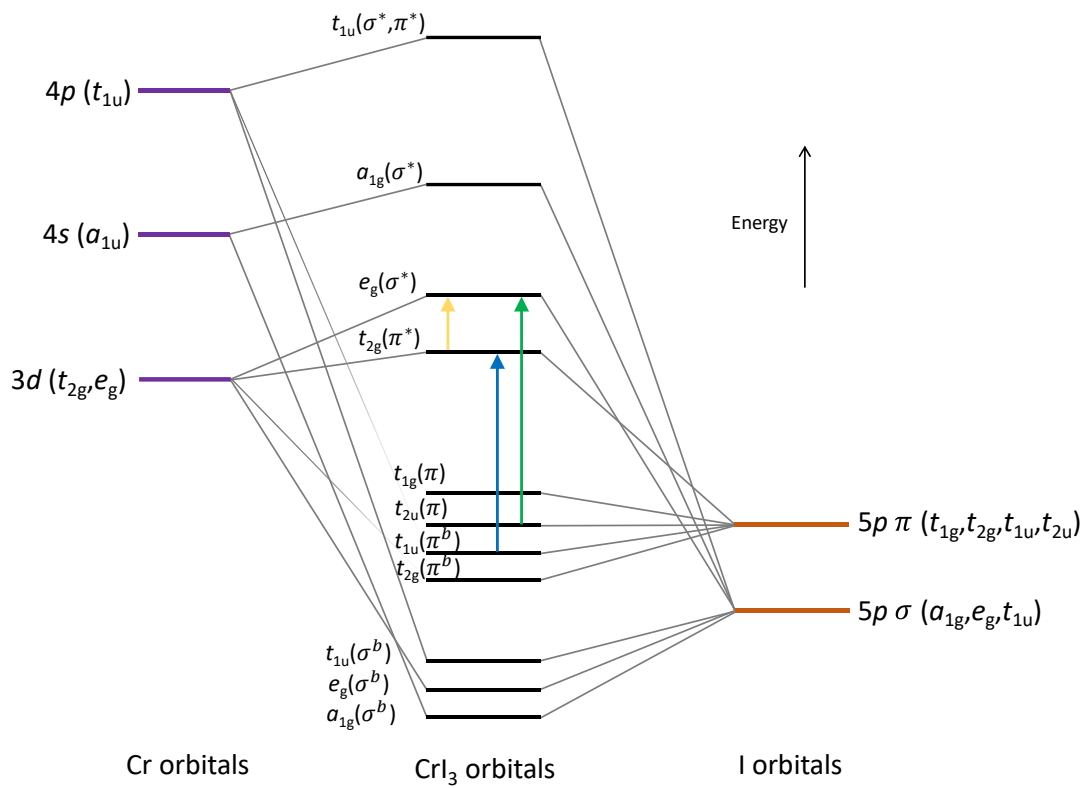

**Fig. S3.** Molecular orbital energy diagram for CrI$_3$. The "b" superscripts and asterisks denote the bonding and anti-bonding orbitals, respectively. The orbitals from $a_{1g}(\sigma^b)$ through $t_{1g}(\pi)$ are completely filled by the iodine electrons. Three spin-aligned electrons occupy $t_{2g}(\pi^*)$ in the ground state. The effect of spin-orbit coupling is not included and the relative energy spacings are not quantitative. The yellow arrow shows a *d-d* transition and the green and blue arrows show two of the possible LMCT transitions. Note that the transition to the $t_{2g}(\pi^*)$ involves an promotion of an electron with spin anti-parallel to the $d^3$ electrons.

Fig. S3 shows a molecular orbital energy diagram for CrI$_3$ (ref. 1,2). The *d-d* ligand-field transitions (Fig. 4b) arise from the ground and excited configurations of the $d^3$ electrons in the $t_{2g}(\pi^*)$ and $e_g(\sigma^*)$ orbitals. Two possibilities for the ligand-to-metal charge transfer (LMCT) transitions are shown in blue and green[3,4], and are described further in Supplementary Section S5. We emphasize that this picture provides a qualitative understanding of the different electronic transitions, and future first-principles calculations will be necessary for quantitative understanding.

**S4. Bulk CrI$_3$ differential reflection spectrum**

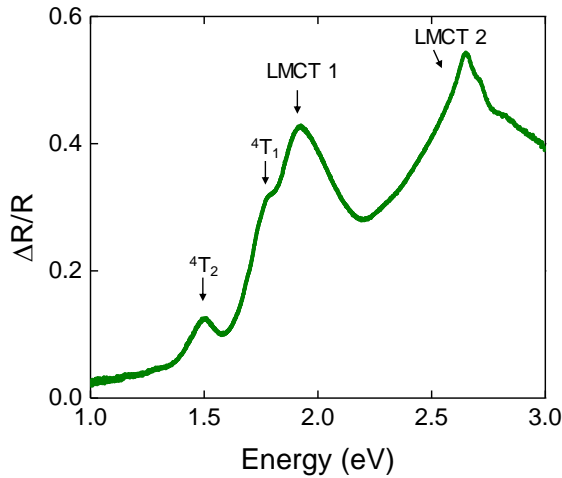

**Fig. S4**. Differential reflection spectrum for thin bulk CrI$_3$.

Figure S4 presents the differential reflection spectrum for thin bulk (~8 layers thick) CrI$_3$ at 15 K. The features are consistent with those reported in Ref. 2. The weak feature at 1.5 eV is the $^4A_2$ to $^4T_2$ transition, which closely matches the monolayer case in Fig. 4a. The stronger features consist of a group of two peaks at 1.80 and 1.93 eV in addition to another set of peaks at 2.65 and 2.70 eV. Based on the connection to prior CrCl$_3$ and CrBr$_3$ experiments[2,5–7] and calculations[3,4], the strong peak at 1.93 eV (LMCT 1) and the 2.7 eV features (LMCT 2) are most likely due to dipole-allowed charge-transfer transitions between the iodine 5$p$ orbitals and Cr$^{3+}$ 3$d$ orbitals. Specifically, the 1.93 eV peak may arise from the $t_{1u}(\pi)$ or $t_{2u}(\pi)$ to $e_g^*$ transition, and the 2.7 eV peaks from the $t_{1u}(\pi)$ or $t_{2u}(\pi)$ to $t_{2g}^*$ (asterisk denotes anti-bonding orbital). We note that there is less confidence in the specific assignment for the higher energy charge transfer feature, although it is accepted as a *p-d* LMCT transition (see ref. 3,4 and references therein). Aside from slight blue-shifts of the peaks in the monolayer limit, we note that the prominent charge-transfer peak at 2.2 eV is not present in the bulk sample, suggesting that interlayer interactions may impact the iodine 5$p$ orbitals, which have significant distribution out of the plane. In addition, the 2.65 eV feature is absent in the monolayer limit, leaving only a single peak at 2.70 eV. Detailed thickness- and polarization-dependent absorption measurements, in addition to first-principles calculations, will be necessary to unravel the intriguing structure of the charge transfer transitions.

## S5. Stokes shift, angular overlap model, and phonon contribution to CrI₃ PL

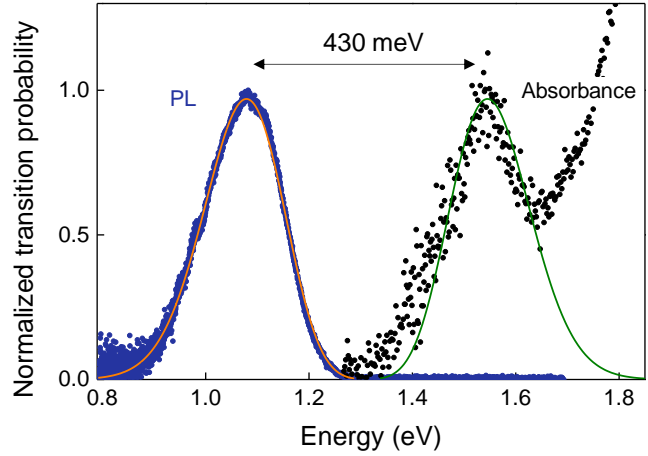

**Fig. S5**. **(a)** Monolayer CrI₃ PL (blue) and absorbance (ΔR/R on sapphire, black), shown with fits described below. The data are from Fig. 1c ($\sigma^-$) and Fig. 4a, at 15 K.

In Fig. S5, we display a fit to the low-temperature monolayer PL spectrum based on the simple, single configurational coordinate model in the harmonic approximation (Fig. 4c) and Franck-Condon principle. The emission at energy $E$ can be approximated by the following[8]:

$$I(E) = \frac{e^{-S}S^p}{p!}\left(1 + S^2 \frac{e^{-\hbar\omega/kT}}{p+1}\right), \quad p = \frac{E_0 - E}{\hbar\omega}$$

Here, $S$ is the Huang-Rhys parameter, $\hbar\omega$ is the effective phonon energy involved in the emission process, $k$ is the Boltzmann constant, $T$ is the temperature, and $E_0$ is the energy of the zero-phonon line. The measured PL and absorbance are first converted to the transition probability by dividing by a factor of $E^3$ and $E$ respectively[8]. Using $T = 15$ K and $E_0 = 1.312$ eV (zero-phonon energy), we achieve a good fit (orange curve in Fig. S5) to the PL spectrum with $S \approx 10$ and $\hbar\omega \approx 24$ meV. The asymmetry and linewidth of the PL spectrum are captured well in this simple model. Because the tail of the 2 eV absorption features overlap with the weak 1.53 eV peak, a careful fit of the absorption is not possible. However, if we plot the corresponding absorption line shape using the same parameters found from the PL fit, we find reasonable overlap with the experimental peak. The similarity between the PL and absorption shapes is further evidence of the shared origin of these optical transitions, namely between the ⁴A₂ ground and ⁴T₂ excited state configurations (shown in Fig. 4c). A large Stokes shift (of 430 meV, in our case) is reasonable for the ⁴T₂ transition in Cr³⁺ complexes[8,9].

The zero-phonon energy for the ⁴T₂ state (~1.3 eV) corresponds to the ligand-field splitting. This energy is determined by the relative strength of $\sigma$ and $\pi$ bonding for Cr-I and can be estimated as $3e_\sigma - 4e_\pi$, where $e_\sigma$ and $e_\pi$ are the angular overlap model parameters for $\sigma$ and $\pi$ bonding[10]. For the Cr-I bonds in an octahedral geometry[11], $e_\sigma = 4100 \; cm^{-1}$ and $e_\pi = 670 \; cm^{-1}$, which gives a ligand-field splitting of ~1.2 eV. The good agreement lends further evidence to the assignment of PL and absorption to the ⁴T₂ state.

The $d$-$d$ transitions become allowed by mixing with odd parity states. For $O_h$ symmetry, the phonon modes that can enable the $d$-$d$ transitions are $t_{1u}$ and $t_{2u}$. The even parity $a_{1g}$, $e_g$, and $t_{2g}$ modes may then contribute to the broadening and structure of the bands[9]. In the Cr trihalides, the trigonal ($D_3$) arrangement of nearest neighbor chromium atoms distorts the octahedral field and thus provide an additional static mechanism through which to intensify the $d$-$d$ transitions, which is similar to what has been suggested for $CrCl_3$ and $CrBr_3$ (ref. 6,7). Finally, we note that future experiments and calculations will be needed to elucidate the details of the phonon modes and intensities, as well as other potentially relevant phenomena, such as excited state Jahn-Teller distortions, spin-orbit coupling, and impurity or substrate broadening.

**References**

1.  Dillon, J. F. & Olson, C. E. Magnetization, resonance, and optical properties of the ferromagnet $CrI_3$. *J. Appl. Phys.* **36,** 1259–1260 (1965).
2.  Grant, P. M. & Street, G. B. Optical properties of the chromium trihalides in the region 1-11 eV. *Bull. Am. Phys. Soc. II* **13,** (1968).
3.  Shinagawa, K., Sato, H., Ross, H. J., McAven, L. F. & Butler, P. H. Charge-transfer transitions in chromium trihalides. *J. Phys. Condens. Matter* **8,** 8457 (1996).
4.  McAven, L. F., Ross, H. J., Shinagawa, K. & Butler, P. H. The Kerr magneto-optic effect in ferromagnetic $CrBr_3$. *J. Phys. B At. Mol. Opt. Phys.* **32,** 563 (1999).
5.  Dillon, J. F., Kamimura, H. & Remeika, J. P. Magneto-optical properties of ferromagnetic chromium trihalides. *J. Phys. Chem. Solids* **27,** 1531–1549 (1966).
6.  Pollini, I. & Spinolo, G. Intrinsic optical properties of $CrCl_3$. *Phys. status solidi* **41,** 691–701 (1970).
7.  Bermudez, V. M. & McClure, D. S. Spectroscopic studies of the two-dimensional magnetic insulators chromium trichloride and chromium tribromide—I. *J. Phys. Chem. Solids* **40,** 129–147 (1979).
8.  Henderson, B. & Imbusch, G. F. *Optical spectroscopy of inorganic solids*. (Clarendon Press, 1989).
9.  Knochenmuss, R., Reber, C., Rajasekharan, M. V & Güdel, H. U. Broadband near-infrared luminescence of $Cr^{+3}$ in the elpasolite lattices $Cs_2NaInCl_6$, $Cs_2NaYCl_6$, and $Cs_2NaYBr_6$. *J. Chem. Phys.* **85,** 4280–4289 (1986).
10. Figgis, B. N. & Hitchman, M. A. *Ligand field theory and its applications*. (Wiley-VCH, 2000).
11. Barton, T. J. & Slade, R. C. Chemical significance of ligand-field parameters in chromium(III) complexes of quadrate symmetry. *J. Chem. Soc. Dalt. Trans.* 650–657 (1975). doi:10.1039/DT9750000650